\documentclass[aps,prd,twocolumn, floatfix, superscriptaddress]{revtex4}

\usepackage{graphicx}
\usepackage{lscape}
\usepackage{indentfirst}
\usepackage{latexsym}
\usepackage{multirow}
\usepackage{tabls}
\usepackage{epsfig}
\usepackage{multirow}

\usepackage{color}
\usepackage{amssymb}

\usepackage{amsfonts}
\usepackage{amsmath}
\usepackage{bm}
\usepackage{mathrsfs}

\usepackage{hyperref}

\usepackage{algorithm}
\usepackage{algpseudocode} 
\usepackage{setspace} 

{

\def\cA{ {\cal A} }

\def\cC{ {\cal C} }

\def\undermui{ \vec{ \underline{\mu}} {}_i } 
\def\undermuj{ \vec{ \underline{\mu}} {}_j } 
\def\underhi{ {\underline{h}_{\undermui}}  }
\def\underhj{ {\underline{h}_{\undermuj}}  }
\def\undermu{ \vec{ \underline{\mu}} }
\def\underh{ {\underline{h}_{\undermu}}  }

\def\MMm{ {\mathrm{MM}} } 
\newcommand{\be}{\begin{equation}}
\newcommand{\ee}{\end{equation}}

\def\lsim{\mathrel{\rlap{\lower3pt\hbox{\hskip1pt$\sim$}}
    \raise1pt\hbox{$<$}}}                
\def\gsim{\mathrel{\rlap{\lower3pt\hbox{\hskip1pt$\sim$}}
    \raise1pt\hbox{$>$}}}         

\def\coordeq{ \, \mathrel{ \rlap{\hbox{\hskip-2.5pt$=$} }
    \raise4pt\hbox{$\cdot$}} \, }                

\begin{document}

\title{Reduced Basis representations of multi-mode black hole ringdown gravitational waves}

\def\addLSU{Department of Physics and Astronomy, Louisiana State University, LA 70803, USA}
\def\addBrownPhys{Department of Physics, Brown University, Providence, RI 02912, USA}
\def\addJPL{Jet Propulsion Laboratory, California Institute of Technology, Pasadena, CA 91125, USA}
\def\addCaltech{Theoretical Astrophysics, California Institute of Technology, Pasadena, CA 91106, USA}
\def\addUMD{Department of Physics, Maryland Center for Fundamental Physics, Joint Space Sciences Institute, Center for Scientific Computation and Mathematical Modeling, University of Maryland, College Park, MD 20742, USA}
\def\addBrownMath{Division of Applied Mathematics, Brown University, Providence, RI 02912, USA}
\def\addUWM{Center for Gravitation and Cosmology, University of Wisconsin-Milwaukee, Milwaukee, WI 53201, USA}
\def\addUMDEO{Department of Physics, Maryland Center for Fundamental Physics, University of Maryland, College Park, MD 20742, USA}

\author{Sarah Caudill}
\affiliation{\addLSU}

\author{Scott E. Field}
\affiliation{\addBrownPhys}
\affiliation{\addUMD}

\author{Chad R.\,Galley}
\affiliation{\addJPL}
\affiliation{\addCaltech} 

\author{Frank Herrmann}
\affiliation{\addUMD}

\author{Manuel Tiglio}
\affiliation{\addUMD}

\begin{abstract}
We construct compact and high accuracy Reduced Basis (RB) representations of single and multiple quasinormal modes (QNMs). The RB method determines a hierarchical and relatively small set of the most relevant waveforms.  We find that the exponential convergence of the method allows for a dramatic compression of template banks used for ringdown searches. Compressing a catalog with a minimal match $\MMm=0.99$, we find that the selected RB waveforms are able to represent {\em any} QNM, including those not in the original bank,  with extremely high accuracy, typically less than $10^{-13}$. We then extend our studies to two-mode QNMs. Inclusion of a second mode is expected to  help with detection, and might make it possible to infer details of the progenitor of the final black hole. We find that the number of RB waveforms needed to represent any two-mode ringdown waveform with the above high accuracy is {\em smaller} than the number of metric-based, one-mode  templates with $\MMm=0.99$. For unconstrained two-modes, which would allow for consistency tests of General Relativity, our high accuracy RB has around $10^4$ {\em fewer} waveforms than the number of metric-based templates for $\MMm=0.99$.  The number of RB elements grows only linearly with the number of multipole modes  versus exponentially with the standard approach, resulting in very compact representations even for many multiple modes. The results of this paper open the possibility of searches of multi-mode ringdown gravitational waves.  
\end{abstract}

\maketitle

\section{Introduction}\label{sec:Intro}

We are quickly approaching the era of gravitational wave astronomy with a world-wide network of advanced interferometric detectors. By 2014, all advanced LIGO interferometers are expected to be operational and will be joined by at least three other detectors, the GEO600, advanced Virgo, and LCGT interferometers. Detection rates for binary neutron star coalescences have been estimated to likely be around $\sim 40\,\mathrm{yr}^{-1}$ and $\sim 20\,\mathrm{yr}^{-1}$ for binary black holes \cite{Abadie:2010cfa,Belczynski:2010tb}. With the potential for discovery and rich science, it will be important to accurately detect these signals. The method of matched filtering, which is currently used for searches of known gravitational waveforms in LIGO's S1-S6 runs, is the optimal method to detect signals buried in Gaussian noise \cite{Kumar2005}.  
While LIGO's noise is not Gaussian, the method, supplemented with additional signal-based vetoes and multi-detector coincidence requirements, is quite suitable for detection purposes, nevertheless. 

Black hole perturbation theory and numerical simulations have shown that, according to General Relativity (GR) and many alternative theories of gravity, perturbed black holes experience an exponentially decaying and oscillatory (or ringdown) phase in which the gravitational wave signal is dominated by a series of quasinormal modes (QNMs) (see \cite{Berti:2009kk} for a review). LIGO ringdown searches  currently  assume that the waveform 
is dominated by the fundamental $l=m=2$ mode \cite{Abbott:2009km}. However, a single mode ringdown search 
for black holes of final mass $M \gtrsim 10^2 M_\odot$ can miss more than 10\% of events in both LIGO and advanced ground-based detectors \cite{Berti:2007zu}.
Furthermore, parameter estimation errors can be large for such single mode searches when the actual waveform contains a second mode \cite{Berti:2007zu}. Two-mode searches have also been proposed for consistency tests of GR and the no-hair theorem \cite{Dreyer:2003bv,Berti:2005ys,Berti:2007zu} and for inferring information about  the progenitors of the final black hole formation \cite{Kamaretsos:2011um}. Detection of the ringdown signal from an intermediate mass black hole may also provide information into the formation history of these potential gravitational wave sources via the black hole's spin \cite{Fregeau:2006yz,Berti:2008af}. Thus, if feasible, it is reasonable to expect that future ringdown searches will look for multi-mode signals.

However, multi-dimensional template banks (or catalogs) can result in such a large number of templates that multi-mode ringdown searches might not be possible in practice.  
The traditional method for building catalogs is to compute a metric for the parameter space that one uses in an algorithm to place points at a given proper distance away from each other and thereby guarantees a minimal match between a signal and template \cite{Owen:1995tm,Creighton:1999pm,Nakano:2003ma,Tsunesada:2004ft,Nakano:2004ib}.
Using this method for a minimal mismatch of 97\%, a two-mode catalog suitable for testing the no-hair theorem with advanced ground-based detectors would require roughly $10^6$ templates versus $10^3$ for a similar one-mode search. This implies a huge increase in the computational cost of matched filtering searches. 

A number of reduced order modeling techniques, such as Proper Orthogonal Decompositions, Singular Value Decompositions, or Principal Component Analysis, can be applied to represent the original template bank by potentially much fewer elements (see, for example, \cite{Heng:2009zz, Brady:2004cf, Cannon:2010qh,Galley:2010rc}).  

Following \cite{Field:2011mf}, in this paper we show that the Reduced Basis (RB) approach provides a dramatically compact representation of multi-mode QNM catalogs. The main two features responsible for such savings are (1) the exponential convergence of the training space representation error (the square of Eq.~(\ref{eq:greedy_error}) below) with respect to the number of  waveforms selected by the algorithm and (2) the linearity of multi-mode ringdown waveforms. 

We summarize some of our key results in this paper. Additional details and an exploration of RB representations for ringdown searches and parameter estimation will be presented elsewhere.

\section{Ringdown waveforms}

Each QNM has a characteristic complex angular frequency $\omega_{\ell mn}$ where $\{ \ell,m \}$ label the angular multipole and $n$ labels the fundamental or overtone index. The real part of $\omega_{\ell m n}$ characterizes the oscillation and equals $2\pi f_{\ell m n}$ where $f_{\ell m n}$ is called the {\it central frequency}. The imaginary part is the inverse of the damping time $ \pi f_{\ell m n} / Q_{\ell m n}$ where $Q_{\ell m n}$ is referred to as the {\it quality factor}. Thus, a single QNM waveform has the form 
\begin{equation}\label{eq:onemodeht}
h_{\ell m n}(t)= \cA_{\ell m n} (\Omega) \frac{M}{r} e^{-\left(\pi f_{\ell m n} / Q_{\ell m n}\right)t} \cos \left(2\pi f_{\ell m n} t\right)
\end{equation}
where $\cA_{\ell m n}(\Omega)$ is the orientation-dependent dimensionless amplitude, $r$ is the distance to the source, $M$ is the black hole mass, and for simplicity we have set the phase and arrival time to zero. We work in units where $G=c=1$ unless otherwise stated. We will ignore the angular dependence in the amplitude and regard $\cA_{\ell m n}$ as constants since, as we discuss in Section \ref{sec:2mode}, the linearity of the problem implies that the reduced bases are valid for all values of the amplitudes and, hence, for all angles.

Defining $h_{\ell m n}(t)=0$ for $t <0$, the Fourier transform of \eqref{eq:onemodeht} is
\begin{equation}\label{onemodehf}
\hat{h}_{\ell m n}(f) \! = \! \frac{ \cA_{\ell m n} (f_{\ell m n} \!-\! 2 i f  Q_{\ell m n}) Q_{\ell m n}}{\pi [ 4 ( f_{\ell m n}^2 \!-\! f^2 ) Q_{\ell m n}^2 + f_{\ell n m}^2 - 4 i f f_{\ell n m} Q_{\ell n m} ]} .
\end{equation}
For our purposes, a multi-mode signal impinging on the detector at time $t=0$ and its Fourier transform are then 
$$
h(t) = \sum_{\ell,m,n} h_{\ell m n}(t) \, , \qquad \hat{h}(f) = \sum_{\ell,m,n} \hat{h}_{\ell m n}(f) \, , 
$$
with $h_{\ell m n}(t)$ and $\hat{h}_{\ell m n}(f)$  given by Eqs.~(\ref{eq:onemodeht}) and (\ref{onemodehf}), respectively. 

The central frequency and quality factor for each multipole can be related to the black hole mass $M$ and dimensionless spin parameter $j$ through the following fitting formulae:
\begin{equation}\label{eq:fitting}
f_{\ell mn} = \frac{f_1 + f_2(1-j)^{f_{3}}}{2\pi M}\, , \quad Q_{\ell mn} = q_1 + q_2(1-j)^{q_3}  
\end{equation}
where values for the $(\ell, m, n)$ dependent constants $f_i$ and $q_i$ can be found in Refs.~\cite{Berti:2005ys,Berti:2007zu}. 
For example, for the $(2,2,0)$ and $(3,3,0)$ modes considered in the following sections we have:
\begin{align} \label{eqn:fit_ell2}
\begin{split}
2\pi M f_{220} & = 1.5251 - 1.1568(1-j)^{0.1292} \\
Q_{220} & = 0.7000 + 1.4187(1-j)^{-0.4990} 
\end{split}
\end{align}
and
\begin{align} \label{eqn:fit_ell3}
\begin{split}
2\pi M f_{330} & = 1.8956 - 1.3043(1-j)^{0.1818} \\
Q_{330} & = 0.900 + 2.3430(1-j)^{-0.4810}.
\end{split}
\end{align}

\section{Reduced Basis} \label{sec:RB}

The RB approach is a framework for efficiently solving parametrized problems, representing the solutions in a compact way, and predicting new ones based on an offline-online decomposition (see \cite{Field:2011mf} and references therein). 
In the current context, the method identifies a set of parameter points such that the associated waveforms constitute a nearly optimal basis for accurate ``spectral" expansions of any other waveform \cite{Binev10convergencerates}.  If the waveforms depend smoothly on the parameters of the problem the expansion is expected to converge very quickly to the original waveform as the number of basis elements is increased. 
We have found that in the case of gravitational waves from non-spinning compact binary inspirals such convergence is exponential \cite{Field:2011mf}. Below we show that this is also the case for ringdown waveforms (see, for example, Fig.~\ref{fig:seeds}). In fact, due to the smooth dependence of the theory on the source parameters we expect the convergence to be exponential for essentially all sources of gravitational waves \footnote{That is, except for boundaries in parameter space delimiting regions with qualitatively different behavior, such as in critical collapse or scattering versus plunge and merger of black holes \cite{Berti:2010ce}, etc.}. 

In what follows we denote a generic waveform, which can be in the time or frequency domain, by $h_{\vec{\mu}}$ and the relevant parameters by $\vec{\mu}$, which is in general multi-dimensional. In the current application, $\vec{\mu}=\{ ( f_{\ell m n },Q_{\ell m n }, \cA_{\ell m n } ) \}$ \footnote{There is actually one less degree of freedom with respect to the number of amplitudes since the waveforms are normalized.}. The scheme starts with a {\it training space} ${\cal T} \equiv \{ \undermui \}_{i=1}^P$ of $P$ numerical values of the parameter $\vec{\mu}$ and associated normalized waveforms $\{ \underhi \}_{i=1}^P$, which we call the {\it training space catalog}. (The underlines denote waveforms or parameters present in the training space.) The natural scalar product  used is directly related to the one of Wiener filtering (see, for example, 
\cite{Maggiore}). For two functions $F,G$ in frequency space
\be
\langle F, G \rangle = \int_{f_L}^{f_U} \frac{F^{\star}(f) G(f)}{S_n(f)} df \, ,  \label{eq:scalar_product}
\ee
with $S_n(f)$ the one-sided power spectral density (PSD) of the detector and $f_L, f_U$ frequency limits implied by the structure of the PSD. The scalar product of (\ref{eq:scalar_product}) is related to the standard overlap integral \cite{Finn:1992wt} by $(F, G) := 4 \, {\rm Re} \langle F, G \rangle$. For definiteness, in this paper we use the Adv LIGO PSD fit of \cite{Ajith:2009fz}. While any PSD may be used, as discussed in \cite{Field:2011mf}, the high accuracy of the reduced basis implies that it can be used with other PSD's with a marginal increase in the representation error. 

The training space can be constructed by any means, including simple random or uniform sampling, more sophisticated stochastic methods \cite{Messenger:2008ta}, the metric approach \cite{Owen:1995tm}, or those of Ref.~\cite{Manca:2009xw}, for example. Regardless of the method used to populate it, the RB formalism produces a compact and highly accurate representation of the training space catalog, 
among other things. 

Part of the output of the algorithm is a sequential selection of $N$ parameter points $\{\vec{\mu}_1,\vec{\mu}_2,\ldots, \vec{\mu}_N  \} \subset {\cal T}$  and their associated waveforms $\{h_{\vec{\mu}_1},h_{\vec{\mu}_2},\ldots, h_{\vec{\mu}_N} \} \subset \{ \underhi \}_{i=1}^P$. The set of waveforms $\{h_{\vec{\mu}_i}\}_{i=1}^N$ (or a linear combination of them) constitutes the {\em reduced basis}. It is sometimes convenient [see Eq.~(\ref{eq:projection})], though not necessary, to work with an orthonormal set $\{ e_i \}_{i=1}^N$ instead of the $\{ h_{\vec{\mu_i}} \}_{i=1}^N$.

The points $\{ \vec{\mu}_i\}_{i=2}^N$ are selected by a {\it greedy algorithm} \cite{Cormen:2001:IA:580470} and are sometimes referred to as {\it greedy points}.  A choice for the first one, $\vec{\mu}_1$ (called the {\it seed}), is needed to initialize the algorithm.  The seed can be chosen arbitrarily, and since the greedy scheme is a global optimization method it is insensitive to its  choice; we explicitly illustrate this in Section \ref{sec:2d} (see Fig.~\ref{fig:seeds}). The algorithm ends when a prescribed tolerance error, defined below in Eq.~(\ref{eq:greedy_error}), in the representation of the training space catalog is reached. For the salient features of the greedy algorithm see Alg.~\ref{alg:Greedy}.

{\scriptsize
\begin{algorithm}[H]
\caption{Brief description of the Greedy Algorithm}
\label{alg:Greedy}
\begin{algorithmic}[1]
\State {\bf Input:} $ \{ \undermui \, , \underhi \}_{i=1}^P$,  $\epsilon$ 
\vskip 10pt
\State {\bf Seed choice} (arbitrary):  $\vec{\mu}_1$
\State RB = $\{ h_{\vec{\mu}_1} \}$
\State $i=1$ and $\epsilon_1 = 1$
\While{$\epsilon_i \ge \epsilon$}
\State $i=i+1$
\State $\epsilon_i = \max_{ \undermu \in {\cal T} }\| P_{(i-1)} \underh - \underh \|$ \label{GreedyErrs}
\State $\vec{\mu}_{i} = \text{argmax}_{ \undermu \in {\cal T} }\| P_{(i-1)} \underh - \underh \|$ 
\State RB = RB $\cup \, h_{\vec{\mu}_i}$
\EndWhile
\vskip 10pt
\State {\bf Output:} RB and greedy points
\end{algorithmic}
\end{algorithm}
}

The reduced basis is used to approximate other waveforms, whether they were in the training space catalog or not, through linear combinations  that represent an orthogonal projection (with respect to the scalar product (\ref{eq:scalar_product})) onto its span, 
\be
h_{\vec{\mu}} = P_Nh_{\vec{\mu}} + \delta h_{\vec{\mu}} \, .  \label{eq:approx}
\ee
The RB approximation is $P_Nh_{\vec{\mu}} $ and satisfies $\langle P_Nh_{\vec{\mu}}, \delta h_{\vec{\mu}} \rangle=0$, by construction. If using an orthonormal basis $\{ e_i \}_{i=1}^N$ then
\be
P_Nh_{\vec{\mu_j}}  = \sum_{i=1}^N \langle e_i, h_{\vec{\mu_j}} \rangle e_i =: \sum_{i=1}^N \alpha_{ij}e_i \, , \label{eq:projection}
\ee
where we have implicitly introduced the projection coefficients $\alpha_{ij}$. 
Another output of the algorithm is precisely the set of $\alpha_{ij}$ projection coefficients for waveforms in the training space catalog, whereas coefficients for any other waveforms may be easily computed.

The greedy algorithm sequentially selects $N=N(\epsilon)$ greedy points and associated waveforms until the {\em maximum} error to represent {\em every} element in the training space catalog reaches the specified tolerance $\epsilon$, 
\be
\epsilon:= \max_{ \undermu \in {\cal T} }\| \delta \underh \| \,,  \label{eq:greedy_error}
\ee
where the norm $\| \cdot \|$ is the one induced by the scalar product (\ref{eq:scalar_product}). 
As discussed in \cite{Field:2011mf}, in the limit of sufficiently dense training spaces the square of $\epsilon$ is comparable to the minimal match ($\MMm$) through \footnote{Strictly speaking, standard template placement algorithms use $1-\MMm$ to characterize how well a discrete template bank captures any signal $h$. On the other hand, $\epsilon^2$ characterizes how well $P_N h$ represents any signal $h$. Albeit different, these error measurements 
are appropriate tools for comparison. }
\be
\epsilon^2 \sim 1 - \MMm {\rm ~~as~~} P \rightarrow \infty. 
\ee
Therefore, we refer to $\epsilon^2$ as the {\it training space representation error} \footnote{Note that in \cite{Field:2011mf} we call $\epsilon$ the {\em greedy error} and here $\epsilon^2$ is called the {\em training space representation error}}. 
The exponential convergence of the RB method implies that $\epsilon$ can be made arbitrarily small with a {\em reduced} number $N$ of {\em basis} elements  (thus the Reduced Basis denomination) with $N < P$ and, in many cases, $N \ll P$. The quantity 
\be
C_r:= P/N \label{eq:compression}
\ee
is called the compression ratio \cite{Salomon2010}. Unless otherwise noted, all the results that we quote are for $\epsilon^2 = 10^{-12}$ in order to avoid roundoff artifacts that gradually appear (otherwise we could choose  $\epsilon^2$ around double precision roundoff, $\epsilon^2 \sim 10^{-14}$). Therefore, any waveform $\underh$ in the training space catalog equals its RB approximation $P_N \underh$ to this level of accuracy.

Another error of interest is associated with the RB representation of {\em any} waveform $h_{\vec{\mu}}$, not necessarily present in the training space catalog. We call this error the {\em waveform representation error}, $\delta (\vec{\mu})$, and define it as, 
\begin{align}
\delta(\vec{\mu}) := || \delta h_{\vec{\mu}} ||^2 & = 1 - {\rm Re} \langle h_{\vec{\mu}} , P_N h_{\vec{\mu}} \rangle 
\label{eq:reperror}
\end{align}
We quantify this error through Monte Carlo simulations,  where we randomly sample 
 $\vec{\mu}$ and compute $\delta(\vec{\mu})$. Remarkably, as discussed in Sections \ref{sec:recovery}  and \ref{sec:3d}, a training space with a modest, finite minimal match will produce a reduced basis that represents the {\em whole space} of QNM waveforms with extremely high accuracy.

The greedy points are hierarchical (i.e. nested) implying that for extra/less accuracy in the RB representation points are simply added/removed. Each greedy sweep is embarrassingly parallel and the computational complexity in going from $i$ to $(i+1)$ basis waveforms is independent of $i$. In other words, the total cost of building a RB with $N$ waveforms is linear in $N$. These features allow the framework to handle large training spaces (cf. Table \ref{tab:Nrb}). Another salient aspect of the RB approach is that it allows one to 
efficiently 
identify on the fly those parameters most relevant for numerical simulations, without any a-priori knowledge of the solutions (see, for example, \cite{Quarteroni}). 

In the simplest, idealized conceptual matched filtering search one would compute the overlap between the signal $s$ and every member of the training space through its RB representation
\be
( s , P_N \underhj ) 
= 4 \, {\rm Re} \sum_{i=1}^N \langle s,e_i \rangle \alpha_{ij} + {\cal O}\left( \epsilon^2 \right) \, .
\ee
 Since the $\alpha_{ij}$ have been precomputed offline, the filtering now involves computing significantly fewer integrals if $N \ll P$, as is usually the case. 
 
This idealized picture gets considerably more involved when including external parameters. The standard
approach for maximizing the filter output over these parameters involves an extremely efficient Fast Fourier Transform (FFT).  However, similar to a Singular Value Decomposition reconstruction, the RB reconstruction computational cost of the same direct FFT strategy would be very high and offset the savings in the reduced order modeling. This is discussed in detail in Ref.~\cite{Cannon:2011tb} so we do not repeat the analysis here since the issue is exactly the same. Strategies such as those proposed in \cite{Cannon:2011tb, Cannon:2011vi} should be investigated in detail to evaluate the efficiency of RB in actual searches.

If one attempted a matched filter search with a RB catalog by filtering each basis function against the data and maximizing over arbitrary linear combinations of these filter outputs, one might encounter high false alarm rates. Instead, it is important to allow only linear combinations that correspond to physical waveforms; see \cite{Field:2011mf} for more details. 

\section{One Mode Ringdown Catalogs} \label{sec:1mode}

\subsection{Metric-based catalogs} \label{sec:metric}

Searches for gravitational waves from perturbed black holes have so far used template banks with only the fundamental $(\ell, m, n) = (2,2,0)$ quasinormal mode (see \cite{Abbott:2009km} and references therein). The ringdown analysis pipeline constructs a 2-dimensional lattice template bank in the $( f,Q)$ space where the mismatch between two templates differing in ringdown frequency by $df$ and in quality factor by $dQ$ is given by the metric \cite{Creighton:1999pm,Abbott:2009km, LSD}
\begin{eqnarray} ds^2 &=& \frac{1}{8} \Bigg[ \frac{3+16Q^4}{Q^2(1+4Q^2)^2} \:
                        dQ^2 - 2 \frac{3+4Q^2}{f Q(1+4Q^2)} \:
                        dQ \: d f \nonumber \\
                      & & \qquad \qquad +\frac{3+8Q^2}{f^2} \: d f^2 \Bigg],
\end{eqnarray}
which assumes white noise.  Currently, the population algorithm uses only the diagonal terms containing $df^2$ and $dQ^2$. Starting with the smallest central frequency and quality factor, the algorithm places templates along $Q$ in steps of
\begin{equation}\label{dQstep}
dQ=\frac{ds_{\mathrm{eff}}Q\left(1+4Q^2\right)}{\sqrt{3+16Q^4}}
\end{equation}
and along $d\phi=d\mathrm{log}(f)$ in steps of
\begin{equation}\label{dlogfstep}
d\phi=\frac{ds_{\mathrm{eff}}}{\sqrt{3+8Q^2}} \, , 
\end{equation}
where $ds_{\mathrm{eff}}=4\sqrt{(1-\MMm)}$ and $\MMm$ is the specified minimal match between template and signal. An example of a ringdown template bank for 
\begin{eqnarray}
\MMm& = &0.99\,, \label{eq:settingsMM} \\
f_{220} &\in & [10,4000]~{\text Hz}\,, \quad  Q_{220} \in [2.1187, 20] \label{eq:settingsfQ}
\end{eqnarray}
is shown in Fig.~\ref{fig:greedy1} as the gray points. This catalog has a total of $2,\!213$ templates. From Eqs.~(\ref{eqn:fit_ell2}) these particular parameter choices correspond to the spin of the black hole $j$ in the range $[0, 0.9947]$ and mass $M$ in the range $[2.9744, 3025.7] M_{\odot}$. While the $(f_{220}, Q_{220})$ parameter space is a rectangle by construction, the corresponding shape in $(M,j)$ space is a ``warped rectangle" and so our stated ranges are not inclusive. Except when otherwise noted, the values and ranges in (\ref{eq:settingsMM}) and (\ref{eq:settingsfQ}) are our default throughout the remainder.

\subsection{One-mode Reduced Basis} \label{sec:2d}

\begin{figure}
\includegraphics[width=\columnwidth]{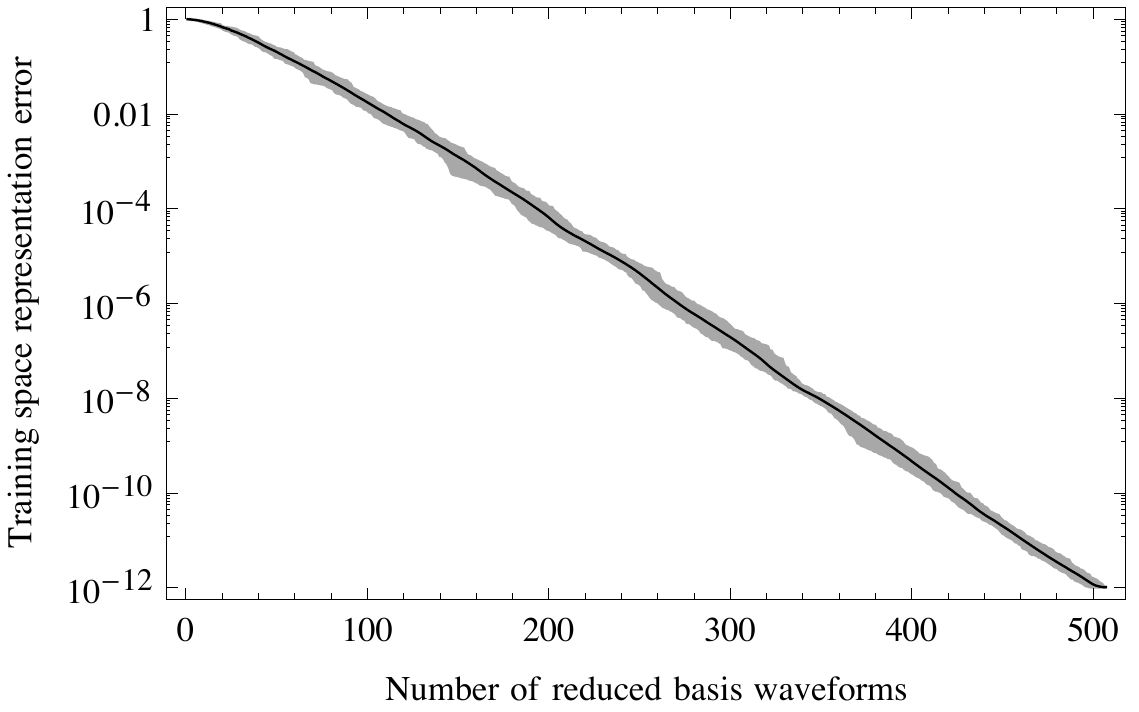}
\caption{Representation error as a function of the number of reduced basis waveforms for a single mode catalog and all possible choices of seeds (see Section~\ref{sec:2d}). The dark line shows the average and the shaded area the maximum dispersion around it. Clearly, the method is robust and does not require any fine tuning. Notice that the exponential rate of convergence of the error is present from the outset.}
\label{fig:seeds}
\end{figure}

\begin{figure}
\includegraphics[width=\columnwidth]{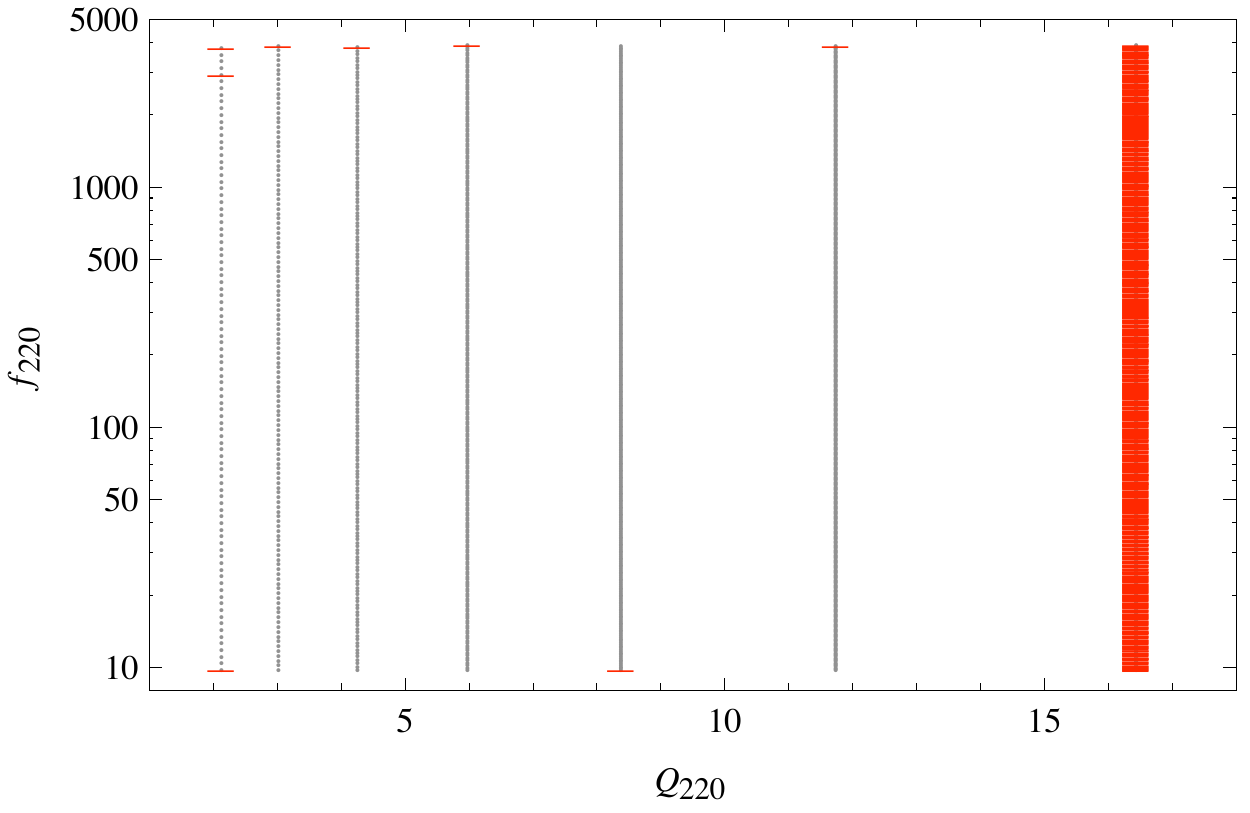}
\caption{One-mode metric placement catalog (gray dots), which is taken to be the training space, and the subset of points selected by the greedy algorithm (red bars) for the $(2,2,0)$ mode. The settings are discussed in Sections~\ref{sec:metric} and \ref{sec:2d}.}
\label{fig:greedy1}
\end{figure}

We begin by using the single-mode QNM metric template bank of Eqs. (\ref{eq:settingsMM},\ref{eq:settingsfQ}) and Fig.~\ref{fig:greedy1} as our default training space for building a reduced basis. In order to explicitly show the robustness of the method we use all $2,\!213$ possible elements of the training space as a seed and run the algorithm $2,\!213$ times, once for each possible seed.
The results are summarized in Fig.~\ref{fig:seeds}, which clearly shows that the  method is robust and  its accuracy and exponential convergence rate do not depend on any fine tuning of the seed. This is actually expected as a consequence of the greedy algorithm being a {\em global} optimization method. As a result, in practice one chooses a single, arbitrary value of the seed.  Notice also from Fig.~\ref{fig:seeds} that the exponential convergence of the error is not asymptotical (i.e. for large number of RB waveforms) but is present from the outset.

Fig.~\ref{fig:greedy1} shows the points in the training space and the subset selected by the greedy algorithm for our default training space representation error of $\epsilon^2 = 10^{-12}$. The selected points are essentially those with the largest quality factor  $Q$ (corresponding to the slowest decaying modes) and a very few extra ones with lower $Q$ and with large and small central frequencies. The figure illustrates the global nature of the algorithm and the redundancy present in sampling via a local criteria. 

As the maximum quality factor in the training space is increased, we find that the number of RB waveforms grows linearly with the corresponding number of templates in the training space, leading to a roughly constant compression ratio [as defined by Eq.~(\ref{eq:compression})] for any chosen $\epsilon$. The qualitative behavior of selected points remains the same, with the vertical right red line in Fig.~\ref{fig:greedy1} shifting to higher $Q$ values and the reduced basis asymptotically resembling a Fourier representation  with a few fast decaying waveforms to account for the damping of QNMs. 

Columns 2 and 3 in Table \ref{tab:Nrb} show the number of RB waveforms needed to represent training space catalogs with different minimal matches. The first element populated by the metric placement algorithm is used as the seed for the greedy algorithm but the results are insensitive to any other choice (cf. Fig.~\ref{fig:seeds} and related discussion).

\begin{center}
\begin{table}
\begin{tabular}{ | c | c | c | c |}
\hline 
\multirow{2}{*}{$1-\MMm$} & \multicolumn{3}{|c|}{1-mode}  \\
	\cline{2-4}
\multirow{2}{*}{~} & ~~$N_{\rm metric}$~~ & $N_{\rm RB} ~(2,2,0)$ & $N_{\rm RB} ~ (3,3,0)$ \\
\hline \hline 
$0.03$ & 999 & 487 & 711   \\
\hline 
$10^{-2}$ & 2,213 & 505 & 732  \\
\hline
$10^{-3}$ &  19,900 & 565 & 930 \\
\hline
$10^{-4}$ &  192,747 & 595 & 972  \\
\hline
$10^{-5}$ &  1,903,689 & 603 & 987  \\
\hline 
\end{tabular}
\caption{Number of RB waveforms ($N_{\rm RB}$) needed to represent 1-mode training spaces for the $(\ell, m, n) = (2,2,0)$ and $(3,3, 0)$  QNMs  with different minimal matches $\MMm$. The training space representation error is taken to be $\epsilon^2 = 10^{-12}$. 
The number of metric-based templates scales with $\MMm$ as $N_{\rm metric} \propto (1-\MMm)^{-1}$ for the 2-dimensional, 1-mode QNM catalog \cite{Owen:1995tm}. }
\label{tab:Nrb}
\end{table}
\end{center}

It might appear from Table \ref{tab:Nrb} that for $\MMm = 0.97-0.99$ and for a high-accuracy training space representation error of $\epsilon^2 = 10^{-12}$ the compression ratio should be modest ($C_r \sim 2-4$) for a single mode catalog and would only be significant for larger $\MMm$s and/or smaller RB representation errors. However, as discussed below in Section~\ref{sec:recovery}, it turns out that the reduced basis represents {\em any} waveform in the given ranges of central frequency and quality factor with extremely high accuracy, leading to huge compression factors even when the reduced basis is built from catalogs with comparatively coarse minimal matches. 
 
As in the case of inspiral waveforms \cite{Field:2011mf}, we find that for any finite range of parameters (e.g., of the central frequency and quality factor) we can represent the {\em whole continuum} of waveforms within any given training space representation error $\epsilon^2 $ by a {\em finite} number of RB waveforms. One way of showing this (as was done in \cite{Field:2011mf}) is by explicitly computing the number of RB waveforms needed to represent training spaces built from different values of $\MMm$. An extrapolation to the limit that the training space becomes the continuum space of waveforms (i.e., $\MMm \to 1$) turns out to asymptote to a finite number of RB waveforms (see also \cite{Cannon:2011xk} where a similar result is found for $\MMm \le 0.99$ in the context of singular value decompositions for inspiral waveforms).  
Another way, which actually has broader implications and has not been demonstrated before, is discussed next.

\subsection{Reduced Basis representation accuracy of arbitrary one-mode waveforms}
\label{sec:recovery}

Every waveform from the training space catalog is represented with an arbitrarily small error ($\epsilon^2 \lesssim 10^{-12}$ in all of our cases) by a relatively small number of RB elements. We now discuss the accuracy of the basis in representing waveforms that are {\it not} necessarily part of the training space. 

Starting from a catalog with a given $\MMm$ to build the basis, the worst possible scenario would be a representation error for some particular waveform $h_{\vec{\mu}}$ of $\| \delta h_{\vec{\mu}} \|^2 \sim (1-\MMm)$. However, one expects the accuracy to be much better since the RB framework exploits the global structure of the template space. As we show below, the accuracy is indeed orders of magnitude better than the pessimistic $1-\MMm$ upper bound.

\begin{figure}
\includegraphics[width=\columnwidth]{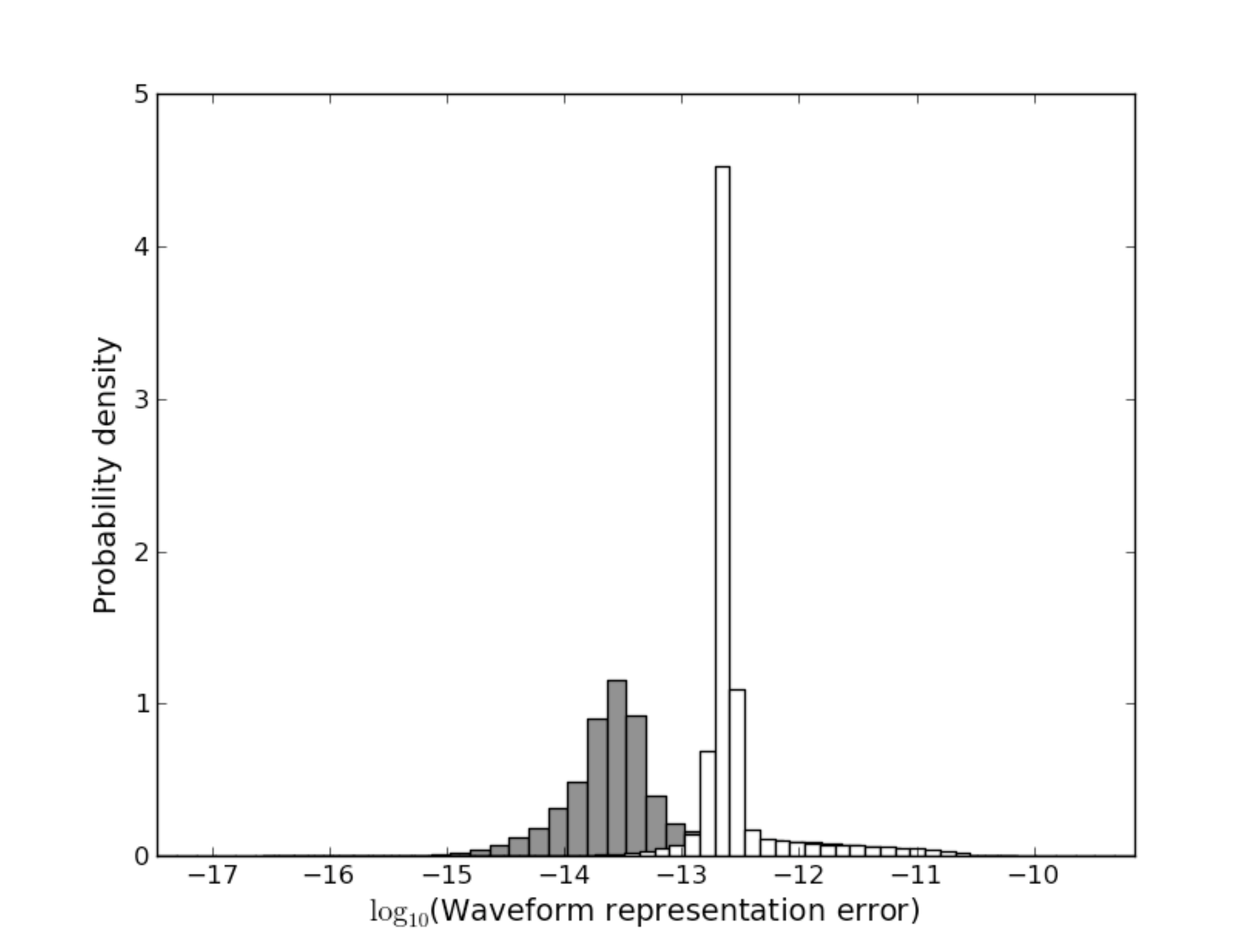} (a) \\
\includegraphics[width=\columnwidth]{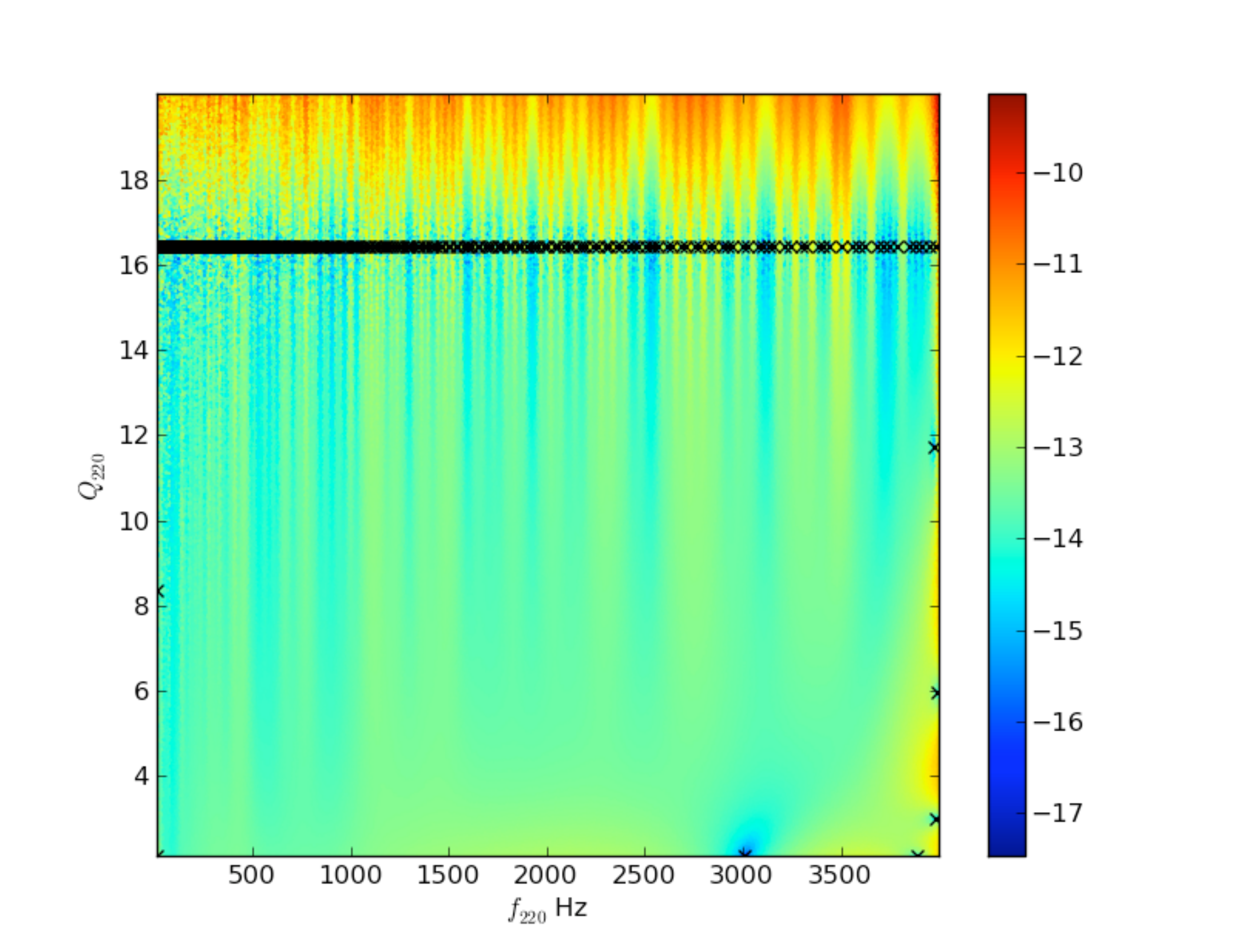} (b)
\caption{Waveform representation error of randomly selected QNMs that are not necessarily present in the training space. (a) Distributions of waveform representation errors for 1-mode (gray) and 2-mode (white) using reduced bases built from a training space with $\MMm=0.99$. In all cases, the error is found to be $< 9\times 10^{-10}$ for single modes and $<9.8\times 10^{-10}$ for two modes. See Section~\ref{sec:recovery} and \ref{sec:3d} for details. (b) The waveform representation error [defined in (\ref{eq:reperror})] as a function of the randomly chosen values of $(f_{220}, Q_{220})$ for the 1-mode case with the same setup as in (a). The color map indicates the value of the error on a $\log_{10}$ scale. 
The worst errors occur for the largest $Q_{220}$ values and are mostly independent of $f_{220}$. The crosses indicate the parameters selected by the greedy algorithm for this reduced basis.}
\label{fig:random}
\end{figure}

We first built a RB representation of our default training space catalog as specified by Eqs.~(\ref{eq:settingsMM}) and (\ref{eq:settingsfQ}). We then randomly sampled more than $6 \times 10^8$ single-mode QNM waveforms drawn uniformly from the same ranges of central frequency and quality factor (\ref{eq:settingsfQ}). We finally evaluated the waveform representation error (\ref{eq:reperror}) for each sample, which gives the fractional signal-to-noise loss from approximating a waveform by its RB representation. The results of these Monte Carlo computations are shown in Fig.~\ref{fig:random}. The error in representing {\em any} of the randomly chosen waveforms with its RB representation is found to be smaller than $9 \times 10^{-10}$. In order to pinpoint this upper bound we refined the Monte Carlo simulations to draw samples from the regions of parameter space with the largest waveform representation errors (i.e., the areas shaded red in Fig.~\ref{fig:random}b). 

The average waveform representation error for a Monte Carlo simulation with $10^7$ points, corresponding to Fig.~\ref{fig:random}, is approximately $4.51 \times 10^{-13}$, which is less than the training space representation error of $\epsilon^2 = 10^{-12}$, and the most frequent value is $\approx 2.6 \times 10^{-14}$.

In other words, in all of our Monte Carlo simulations, {\em any} waveform, not just those present in the training space catalog, is found to be represented by the RB with extremely high accuracy. 

A simple extrapolation from Table~\ref{tab:Nrb} shows that $\sim 10^{10}$ metric templates would be needed to achieve a $\MMm$ comparable to the above waveform representation error of $\sim 10^{-9}$ for our reduced basis with $505$ elements, implying an effective compression ratio of $C_r \sim 10^7$. Furthermore, the fact that our simulations strongly indicate that the maximum waveform representation error is strictly bounded (by $9\times 10^{-10}$ in the settings discussed) would imply a formally infinite compression ratio. This again reflects another facet of being able to represent the whole spectrum of waveforms with a finite number of basis elements. The key distinction here, compared to the discussion at the end of the previous subsection (\ref{sec:2d}), is that one does not need to explicitly compute a set of bases with increasingly larger number of training space points to reach a target accuracy limit. This would allow for the \emph{a priori} construction of high accuracy reduced bases for sources with larger number of parameters, such as precessing binary inspirals, at a fraction of the projected computational cost. 

We have done exhaustive additional tests to support the conclusion that the representation error is extremely small for {\em all} waveforms with parameter values in the set ranges [e.g., as in Eq.~(\ref{eq:settingsfQ})]. For example, since a random, uniform sampling of $(f_{220} ,Q_{220} )$ leads to a non-uniform sampling of $(M,j)$ [see Eq.~(\ref{eqn:fit_ell2})]
we have performed Monte Carlo simulations with random, uniform sampling of mass and spin such that the corresponding central frequencies and quality factors are within the ranges of Eq.~(\ref{eq:settingsfQ}). In addition, instead of selecting waveforms at random, we have used all those from a metric-based catalog with a  minimal match of about $0.9999$, corresponding to more than $10^5$ templates. We represented the waveforms using the RB generated by our default training space [from Eqs.~(\ref{eq:settingsMM}) and (\ref{eq:settingsfQ})] and computed the corresponding waveform representation errors.
In all cases the errors were found to be bounded by the previously quoted value of $9\times 10^{-10}$.

As discussed in the next section, the high accuracy associated with representing arbitrary waveforms by a RB  carries over to the case of multiple QNMs.

\section{Multi-Mode Ringdown catalogs} \label{sec:2mode}

In this section we discuss how the RB approach can be used to efficiently build, compress, and represent the space of multiple QNM waveforms. We present two kinds of multi-QNM reduced basis. The first is for a two-mode ringdown waveform wherein the two central frequencies and quality factors are related to each other through a constraining relation provided by GR for distorted black holes. 
The second is for a ringdown waveform consisting of $p$ modes where the central frequencies and quality factors are unconstrained by any model or theory. In particular, this latter case is amenable for providing a test of the no-hair theorem \cite{Berti:2007zu}, through a test of the precise relationship between the multiple ringdown modes as predicted by GR.

\subsection{Constrained two-mode Reduced Basis} \label{sec:3d}

Here we consider two-mode ringdown waveforms with $(\ell, m, n) = (2,2,0)$ and $(3,3,0)$, which are of the form 
\be
h = {\cal C} \left[ (1-\cA)h_{220}  + \cA h_{330}\right]   \label{eq:2mode}
\ee
where $\cA \in [0,1]$ and $\cC$ is fixed by the normalization condition $\langle h , h \rangle=1$. 

The parameters for this two-mode QNM are $(f_{220}, Q_{220})$, $(f_{330}, Q_{330})$, and the relative amplitude parameter $\cA$. However, GR provides a relation between $(f_{\ell m n}, Q_{\ell m n})$ and the mass and spin of a perturbed black hole. From Eqs.~(\ref{eq:fitting}) or, specifically for the two-mode case here, Eqs.~(\ref{eqn:fit_ell2}) and (\ref{eqn:fit_ell3}) it follows that $(f_{330}, Q_{330})$ are related to $(f_{220}, Q_{220})$ so that there are only three independent parameters. In this sense, GR constrains the otherwise 5-dimensional parameter space to a 3-dimensional one, namely, $\{ f_{220}, Q_{220}, \cA\}$ (or, equivalently, $\{ j, M, \cA\}$). 

For a given $\MMm$ and for $\cA \in [0,1]$ we build a training space for the $(2,2,0)$ mode using the metric approach as described in Section~\ref{sec:metric} \footnote{One could use any method to populate the training space, as mentioned in Section \ref{sec:RB}, but we use the metric approach since it is available for this case and the training space then has a well-defined interpretation in terms of minimal matches.}.  
Next, using the fitting formulae given by Eq.~(\ref{eqn:fit_ell2}), the values of mass and spin for each template are determined. Then, using Eq.~(\ref{eqn:fit_ell3}),  the corresponding values of $(f_{330}, Q_{330})$ are computed and paired to the starting $(f_{220}, Q_{220})$. This procedure is repeated for each template in the $(2,2,0)$ catalog and subsequently populates the $(f_{330}, Q_{330})$ plane with templates that are ``inherited'' from the $(2,2,0)$ training space. The relative amplitude parameter $\cA$ is sampled with $n_{\cA}$ equally spaced points.  In this section and in Table~\ref{tab:Nrb} for the $(3,3,0)$ mode column, $\MMm$ refers to the minimal match of the  $(2,2,0)$ starting catalog.

\subsubsection{Results}

\begin{figure}
\includegraphics[width=\columnwidth]{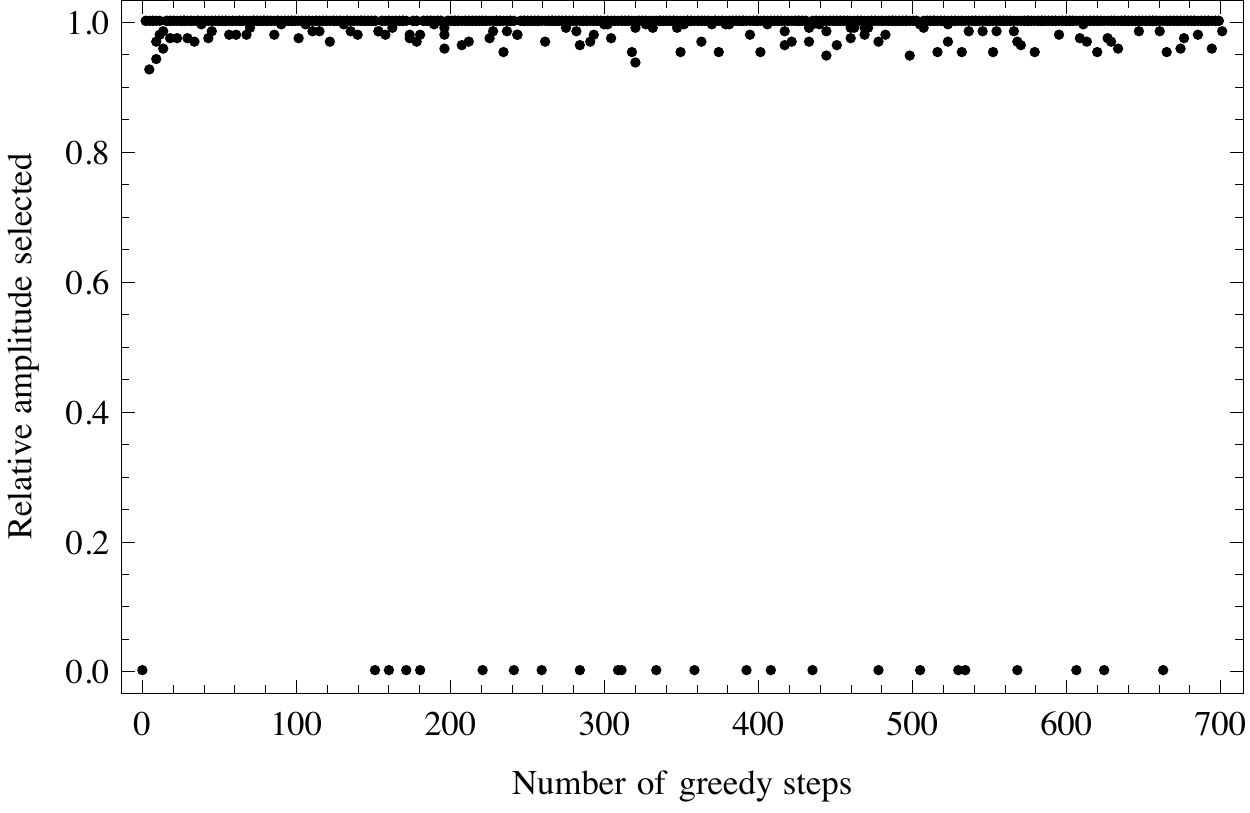}
\caption{Values for the relative amplitude parameter $\cA$ in the two-mode constrained waveforms [Eq.~(\ref{eq:2mode})] selected by the greedy algorithm for each reduced basis waveform (indexed on the horizontal axis). Here there are $n_{\cA}=1,000$ samples for $\cA\in[0,1]$ but essentially either the $(2,2,0)$ mode ($\cA=0$) {\em or} the $(3,3,0)$ mode ($\cA=1$) are selected, with a higher density of points associated with the latter due to its larger maximum quality factor for any given black hole spin parameter value. 
About $3\%$ of the selected amplitudes are {\em equal} to zero (not just nearly so). All of the remaining points selected are either {\em equal} to 1 (about $79\%$) or between $\cA=0.93$ and 1 (about $18\%$).} 
\label{fig:amp3d}
\end{figure}

\begin{figure}
\includegraphics[width=\columnwidth]{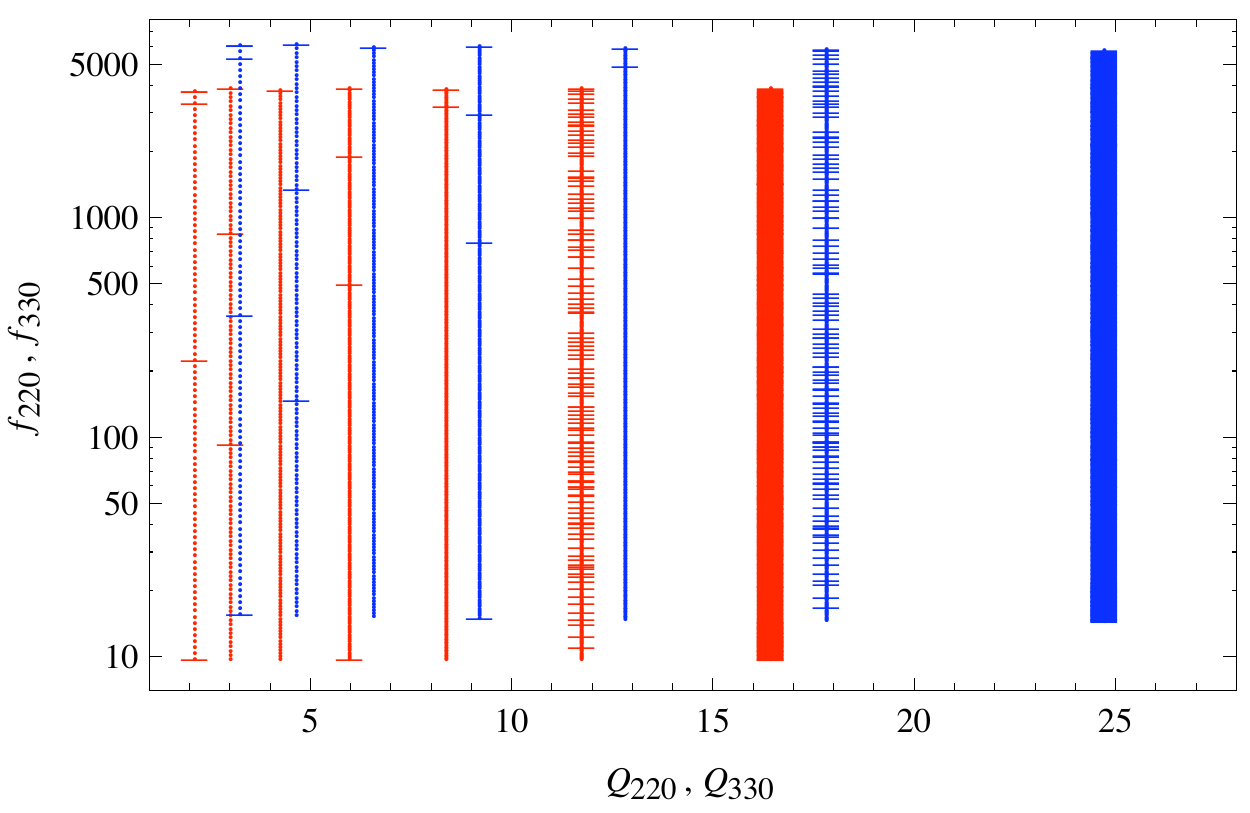}
\caption{The metric-based training space (points) for the same constrained, two-mode catalog of Fig.~\ref{fig:amp3d} and the subset of parameter values selected by the greedy algorithm (bars) for a training space representation error of $\epsilon^2 = 10^{-12}$. Red: $(2,2,0)$ mode. Blue: $(3,3,0)$ mode. The distribution of points closely resembles a subset of the two one-mode greedy points due to the linearity of the problem and  to the ``decoupling'' of selected amplitude points (see Fig.~\ref{fig:amp3d}).}
\label{fig:greedy2}
\end{figure}

In all of our numerical simulations we have found that, for any fixed training space representation error $\epsilon^2$,  the number of reduced basis waveforms is {\em independent of $n_{\cA}$ for $n_{\cA}\geq 2$}. This is not too surprising, given the linearity of the two-mode waveform (\ref{eq:2mode}); see also the discussion in Section~\ref{sec:5d}. A related observation is that, as shown in Fig.~\ref{fig:amp3d}, the greedy algorithm essentially selects {\em either} the $(2,2,0)$ mode ($\cA=0$) {\em or}  the $(3,3,0)$ one ($\cA=1$) as opposed to combinations of them. As a result, the distribution of parameter points selected by the  algorithm closely resembles the union of a subset of points for two individual, one-mode reduced bases. This is illustrated in Figure~\ref{fig:greedy2}, which should be compared to Figure~\ref{fig:greedy1} for a one-mode $(2,2,0)$ reduced basis.

The number of constrained, two-mode RB waveforms $N_{\rm RB}$ for different values of $\MMm$  is shown in Table~\ref{tab:Nrb2mode} in the column labeled ``2-mode, GR.''
Notice that these values are only marginally larger than those corresponding to a one-mode $(3,3,0)$ basis in Table~\ref{tab:Nrb}. To understand why, notice from Fig.~\ref{fig:amp3d} that the $(3,3,0)$ mode, corresponding to $\cA=1$ in (\ref{eq:2mode}), is chosen much more frequently than the $(2,2,0)$ mode. In turn, the greedy algorithm's preference for $\cA=1$ can be understood by the fact that: i) for any maximum value of $Q$ in the starting $(2,2,0)$ catalog the corresponding value for the $(3,3,0$) mode for the same black hole spin parameter $j$ is larger (see Eqs.~(\ref{eqn:fit_ell2}) and (\ref{eqn:fit_ell3})), and ii)  as we have already discussed in Section~\ref{sec:3d}, the greedy algorithm for a single-mode RB mostly selects points with the largest value of $Q$ in the catalog (see Figure~\ref{fig:greedy1}).

Table~\ref{tab:Nrb2mode} also shows the approximate number of templates $N_{\rm metric}$ for constrained two-modes using the metric placement method in the column labeled ``2-mode, GR.'' 
The compression ratio with respect to the metric number of templates for $\MMm = 0.99$ is  $C_r \approx 24$ and dramatically increases as $\MMm$ increases. The number of RB waveforms asymptotically approaches about $1,\!000$, which suggests that a two-mode search with the RB approach may be easily feasible.  

Next, we show that the accuracy of 2-mode reduced bases and the associated compression ratios are much better, like in the 1-mode case of Section \ref{sec:recovery}, than initially indicated in Table~\ref{tab:Nrb2mode}, which are already very good.

\begin{center}
\begin{table}
\begin{tabular}{ | c | c | c | c | c |}
\hline 
\multirow{2}{*}{$1-\MMm$}  & \multicolumn{2}{|c|}{2-mode, GR} &\multicolumn{2}{|c|}{2-mode} \\
	\cline{2-5}
\multirow{2}{*}{~}  & ~~ $N_{\rm metric}$ ~~ & $~~N_{\rm RB}~~$ & ~~$N_{\rm metric}$~~ & $~~N_{\rm RB}~~$ \\
\hline \hline 
$0.03$  & $3.5 \times 10^{3}$  & 737 & $3.4 \times 10^{6}$ & 1,198   \\
\hline 
$10^{-2}$ & $1.8 \times 10^{4}$ & 751 &  $5.3 \times 10^7$ &1,237  \\
\hline
$10^{-3}$ & $5.8 \times 10^{5}$ & 958 &  $1.9 \times 10^{10}$ & 1,495 \\
\hline
$10^{-4}$ & $1.8 \times 10^{7}$ & 1,007 &  $5.3 \times 10^{12}$ & 1,567  \\
\hline
$10^{-5}$ & $5.8 \times 10^{8}$ & 1,018 & $1.9 \times 10^{15}$ & 1,590  \\
\hline 
\end{tabular}
\caption{Number of reduced basis waveforms ($N_{\rm RB}$) needed to represent 2-mode training spaces with $(\ell, m, n) = (2,2,0)$ and $(3,3, 0)$ for  different minimal matches $\MMm$. The training space representation error is $\epsilon^2 = 10^{-12}$.  
The ``2-mode, GR'' and ``2-mode'' cases are discussed in Sections~\ref{sec:3d} and \ref{sec:5d}, respectively.  
The number of metric-based templates in those two cases scales with $\MMm$ as $N_{\rm metric} \propto (1-\MMm)^{-d/2}$, with $d=3$ and $d=5$, respectively \cite{Owen:1995tm}. 
For the ``2-mode, GR'' entries, $\MMm$ refers to the minimal match associated with the training space for the $(2,2,0)$ mode, which is used to generate the points in the $(f_{330}, Q_{330})$ plane as described in the text.
For the ``2-mode'' case the training space representation error of $10^{-12}$ corresponds to each mode separately and the corresponding total error is bounded from above by $4 \times 10^{-12}$, as discussed in Section \ref{sec:5d}. We are grateful to V.~Cardoso for his help in generating the values of $N_{\rm metric}$ in the second column for the constrained, 2-mode case.}
\label{tab:Nrb2mode}
\end{table}
\end{center}

\subsubsection{Reduced Basis representation accuracy of arbitrary two-mode waveforms}

In analogy with the 1-mode case, we have found that two-mode, constrained reduced bases built from relatively coarse training spaces turn out to represent {\em any} waveform with extremely high accuracy. Since the details follow those of Section~\ref{sec:recovery} our description here is more succinct. 

In order to emphasize our finding that for constrained, two-mode ringdown waveforms the number of basis elements saturates at $n_{\cA}=2$ amplitude samples, we describe waveform representation error results for a reduced basis built with exactly $n_{\cA}=2$. That is, from here on the training space and, as a consequence, the resulting reduced basis only includes  $\cA=0$ and $\cA=1$ relative amplitudes in the waveforms. 

We have uniformly sampled the constrained, 2-mode parameter space $(f_{220},Q_{220},\cA)$ with $f_{220}$ and $Q_{220}$ in the same range used to build the reduced basis, Eq.~(\ref{eq:settingsfQ}), and with $\cA \in [0,1]$. The corresponding values of $f_{330}$ and $Q_{330}$ were computed as described earlier using the constraint from General Relativity that they correspond to the same black hole mass and spin as those of $(f_{220},Q_{220})$. The resulting waveforms were then projected onto the reduced basis built with a training space of $\MMm=0.99$ as described above and the waveform representation error (\ref{eq:reperror}) for each sample was computed.  We have executed many Monte Carlo simulations, with the largest one having $10^9$ random triples $(f_{220}, Q_{220}, \cA)$. In all cases the maximum waveform representation error was found to be below $9.8 \times 10^{-10}$, which is remarkably close to our strict numerical upper bound found for the one mode case described in Section~\ref{sec:recovery}, namely, $9 \times 10^{-10}$. As an example, Fig.~\ref{fig:random}a shows a histogram of the waveform representation error for a Monte Carlo simulation with $10^7$ points. In this case, the average error is about $7.35 \times 10^{-13}$, which is less than the training space representation error of $\epsilon^2 = 10^{-12}$, and the most frequent value is $\approx 2.4 \times 10^{-13}$.

\begin{figure}
\includegraphics[width=\columnwidth]{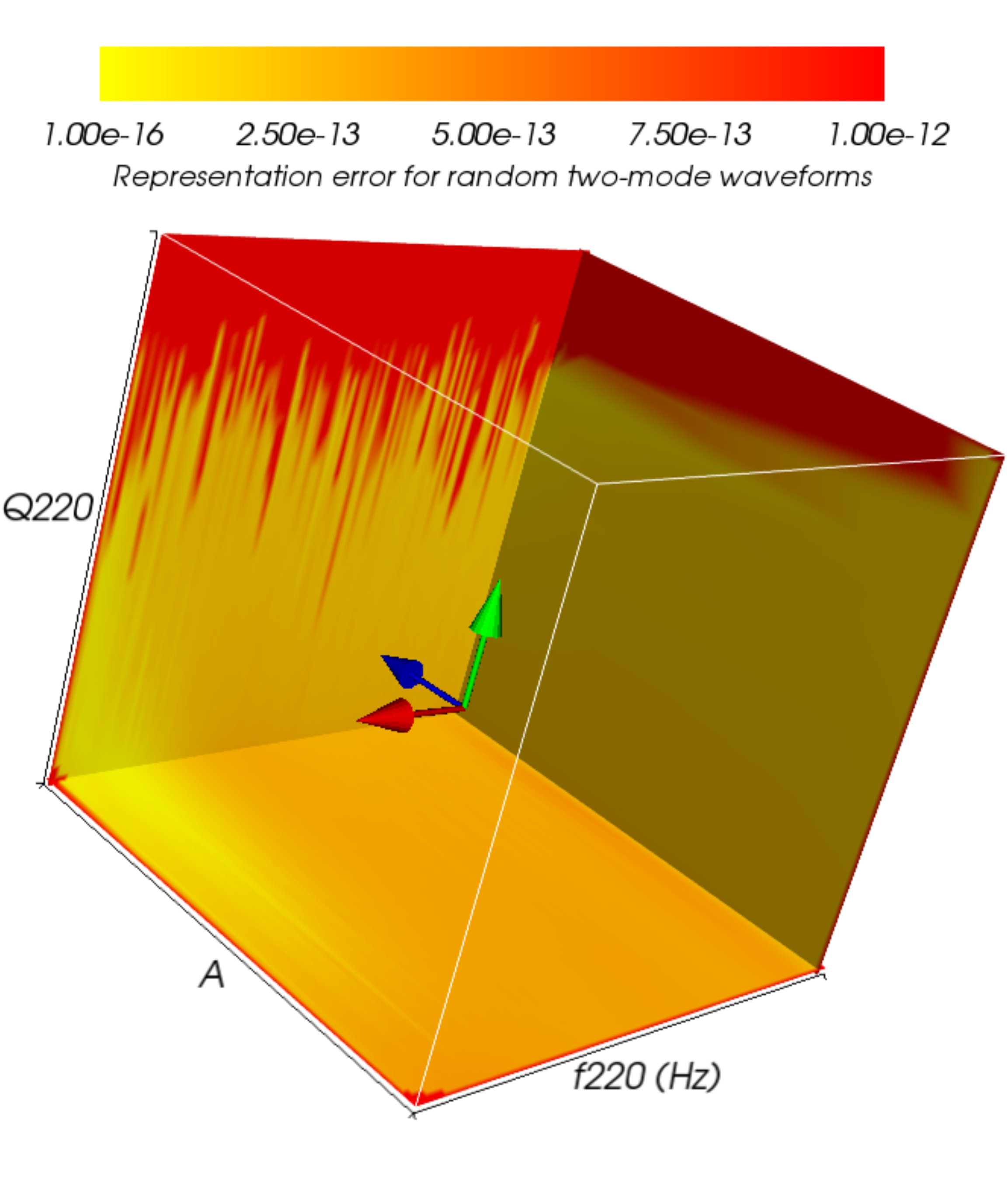}
\caption{Waveform representation errors, as defined by Eq.(\ref{eq:reperror}), of constrained, two-mode waveforms for a random sample of $(f_{220},Q_{220},\cA)$ triples. The training space used to build the reduced basis is relatively coarse yet the waveform representation error is found to be, in all cases, smaller than $9.8 \times 10^{-10}$. See Section \ref{sec:3d} for more details.}  
\label{fig:3d_random}
\end{figure}

Fig.~\ref{fig:3d_random} shows the distribution of the waveform representation error as a function of the sample values for $(f_{220}, Q_{220}, \cA)$. As with the 1-mode case, the largest errors occur for larger $Q_{220}$ values and are mostly independent of $f_{220}$ and the relative amplitude.

\subsection{Unconstrained multi-mode Reduced Basis} \label{sec:5d}

As mentioned in Section~\ref{sec:Intro}, searches for multi-mode ringdown gravitational waves have been proposed as a consistency test of GR and, more specifically, the no-hair theorem. The basic idea is not to enforce the constraint that the ringdown frequencies of the different modes should correspond to the same black hole mass and spin, as in the previous subsection (\ref{sec:3d}), but to check it \emph{a posteriori}, by assuming the standard relationship in Eq.~(\ref{eq:fitting}) between QNM frequencies and black hole mass and spin for {\em each} mode. 

The dimension of the parameter space for $p$ unconstrained fundamental modes \footnote{As it should become clear, this is just a simplifying assumption and the general argument does not rely on it.} is $(3p-1)$: $2p$ for $(f_{lmn},Q_{lmn})$ and $(p-1)$ for the relative amplitudes of the normalized waveform. Already for $p=2$ this results in a very large bank of  roughly $10^6-10^7$ templates when using the metric approach
for $\MMm = 0.97-0.99$ (see Table \ref{tab:Nrb2mode}). 

One could build a ``full'' reduced basis from a $(3p-1)$-dimensional training space.
However, the linearity of the problem and the results of Section~\ref{sec:3d} suggest a much simpler approach.  Since ringdown waveforms are a {\em linear} superposition of single modes [in the following the index $I$ generically labels the  $(\ell,m,n)$ triple], 
\be
h= \sum_{I=1}^p \cA_I h_I \, ,  \label{eq:pmode}
\ee
then a ``simple" RB representation for (\ref{eq:pmode}) could be given by building a reduced basis for each $I$-th mode separately from the $I$-th training space ${\cal T}_I = \{ \vec{\underline{\mu}}{}_{Ii} \}_{i=1}^{P_I}$ and to define the representation of the multi-mode wave to be the sum of the individual projections 
\be
P_N h_{\vec{\mu}} := \sum_{I=1}^p \cA_I P_{N_I} h_{I, {\vec{\mu}}} \label{eq:RB_pmode} \, . 
\ee
Here $P_{N_I}$ denotes the standard orthogonal projection onto the reduced basis for the single $I$-th mode built from a training space catalog associated with ${\cal T}_I$. Denoting by $\epsilon_I^2$ the training space representation error for the $I$-th mode,
\be
\epsilon_I^2 := \max_{ \vec{\underline{\mu}}{}_I \in {\cal T}_I} \| \underline{h}_{I, \vec{\underline{\mu}}{}_I} - P_{N_I} \underline{h}_{I, \vec{\underline{\mu}}{}_I}  \|^2 \, ,  \label{eq:i-error}
\ee
then the total training space representation error for (\ref{eq:pmode}) is bounded by 
$$
\epsilon_{\rm simple}^2:= \max_{ \vec{\underline{\mu}} \in {\cal T} } \| \underline{h}_{\undermu} - P_{N} \underline{h}_{ \undermu}  \| ^2 \leq \sum_{I=1}^p \epsilon_I^2 +  \sum_{\substack{I,J=1 \\ I\neq J}}^ p \epsilon_I \epsilon_J \, ,  
$$
where the full training space ${\cal T}$ is the product of the individual training spaces $\bigotimes_{I=1}^p {\cal T}_I$. 
If all the errors are chosen to be comparable (i.e, $\epsilon_I \sim \epsilon$) then the bound is simply
\be
\epsilon_{\rm simple}^2 \lesssim p^2 \epsilon^2\,. \label{eq:error_nmode}
\ee
In most cases of practical interest for gravitational wave searches, including advanced and third generation earth-based detectors such as the Einstein telescope,
$p\leq 4$ \cite{Kamaretsos:2011um}.  If $\epsilon^2 = 10^{-12}$, as in all of our cases, then the total training space representation error would still be very small, of order $\epsilon_{\rm simple}^2 \lesssim {\cal O}\left( 10^{-11} \right)$. In fact, one could include up to $p=10^5$ modes and still have a maximum total error $\epsilon_{\rm simple}^2 \lesssim 0.01$, which is comparable in terms of waveform representation accuracy to a minimal match of about 0.99. This feature is another advantage of having RB representations with very high accuracy, even if the latter exceeds the accuracy of the detector itself or of the physical modeling. Such representations leave room for simplifications while preserving high accuracy in the overall approach. 

The total number $N= \sum_{I=1}^p N_I$ of RB waveforms scales {\em linearly} with $p$. If $N_I \sim {\cal N}$ basis elements are needed to represent each mode   
 then $N\sim p {\cal N}$. In contrast, the number of metric-based templates is proportional to $(1-\MMm)^{-(3p-1)/2}$ \cite{Owen:1995tm} and increases dramatically with $p$ for any fixed $\MMm$ (see, for example, Tables \ref{tab:Nrb} and \ref{tab:Nrb2mode}). 

Furthermore, the representation (\ref{eq:RB_pmode}) is valid for all values and ranges of amplitudes $\cA_I$. 
In particular, $N$ is independent of the range or values of the $\{ \cA_I \}$. This last observation, though trivial, helps to understand the results of  Section~\ref{sec:3d} for constrained 2-mode waveforms: essentially one mode {\em or} the other is selected at each step by the greedy algorithm (see Fig.~\ref{fig:amp3d}), the number of RB waveforms saturates at $n_{\cA}=2$ samples ($\cA=0$ and $\cA=1$), and the selected $(f,Q)$ points (Fig.~\ref{fig:greedy2}) resemble a subset of those picked by the two 1-mode reduced bases. 

Already for the $p=2$ case and template banks with minimal matches as low as $\MMm=0.97$ and $0.99$, this simple RB approach for unconstrained multi-mode QNMs implies a compression ratio of around {\em three and four orders of magnitude relative to the respective training spaces}. In addition, as discussed in Sections \ref{sec:recovery} and \ref{sec:3d}, the resulting bases actually represent the {\em whole space of waveforms} with an accuracy comparable to the  training space representation error, resulting in effectively much larger, if not formally infinite, compression ratios. The number of unconstrained 2-mode [$(2,2,0)$ and $(3,3,0)$] metric-based templates for different values of $\MMm$ and the number of RB waveforms needed to represent them with this simple approach are given in Table \ref{tab:Nrb2mode}.

\section{Final Remarks}

We have shown that the RB approach provides very compact and high-accuracy representations of multi-mode ringdown gravitational waves. For example, the number of RB waveforms needed to represent {\em any two-mode} General Relativity quasinormal mode 
with 
a representation error of $10^{-12}$ is {\em smaller} than the number of metric-based templates for a {\em one-mode} catalog with a minimal match of $\MMm \sim 0.99$. The comparison with the number of two-mode, metric-based templates for a catalog with $\MMm \sim 0.97-0.99$ designed to test the consistency of GR and the no-hair theorem is even more striking: the number of RB waveforms needed to represent the continuum is $3$--$4$ orders of magnitude smaller than the number of metric-placed templates for minimal matches of $0.97$--$0.99$.

We have demonstrated the robustness of the RB method by observing essentially identical training space representation errors regardless of the seed value. In the case of one-mode ringdown waveforms, we 
showed how the greedy algorithm preferentially selects waveforms with the largest quality factors to construct the basis. The number of RB waveforms for the $(\ell,m,n)=(2,2,0)$ one-mode case needed to represent the continuum asymptotes to about $600$ for advanced ground-based gravitational wave detectors. 
 We observed that the compression ratio remains constant as a function of the maximum quality factor. Through detailed numerical studies we demonstrated that any waveform that falls within the one- and two-mode training space's ranges in central frequency and quality factor (and, in the case of 2-modes, for an arbitrary relative amplitude) can be recovered with an error no larger than $\approx 9\times 10^{-10}$. Thus, we can construct high accuracy reduced basis representations of these ringdown waveforms very easily off-line. We expect this result to extend to RB representations of spinning binary inspirals and other multi-dimensional systems that depend smoothly on their parameters.

In the case of constrained two-mode ringdown waveforms, we find that the RB method, in additional to its exponential convergence, benefits from the linearity of the QNMs. The number of RB waveforms is independent of the number of relative amplitude parameter samples $n_{\cA}$ for $n_{\cA}\geq 2$ for any fixed training space representation error. The greedy algorithm essentially only selects, for example,  the (2,2,0) mode or the (3,3,0) mode. The number of RB waveforms for the constrained two-mode case to represent the continuum asymptotes to about $1,000$ for advanced ground-based detectors. Finally, in the case of unconstrained two-mode ringdown waveforms, we again exploit the linearity of the problem to propose the construction of a ``simple'' RB for each mode separately. Then the representation of the multi-mode wave is simply the sum of the individual projections. In addition, the number of RB waveforms scales linearly with the number of modes indicating a very large reduction in the computational cost of multi-mode ringdown searches.

The results of this paper open up the possibility of searches of multi-mode ringdown gravitational waves thereby allowing one to test the no-hair theorem, to significantly reduce the event loss rate of these kinds of signals, to improve parameter estimation, and to possibly infer progenitors from the relative amplitudes of the different QNMs. These results are made freely available~\cite{RBM_Website}. 

\section{Acknowledgments}
We are especially grateful to Emanuele Berti and Vitor Cardoso for very valuable discussions (which motivated this work), feedback, assistance, and suggestions. We also thank Evan Ochsner, Duncan Brown, Nickolas Fotopoulos, Ajith Parameswaran, Ulrich Sperhake and Bernard Whiting for very useful discussions and feedback on the manuscript.
This work has been supported by NSF Grants NSF PHY-0905184 to Louisiana State University, PHY1005632 and PHY0801213 to the University of Maryland, and NSF DMS 0554377 and OSD/AFOSR FA9550-09-1-0613 to Brown University. C.R.G. has been supported by an appointment to the NASA Postdoctoral Program at the Jet Propulsion Laboratory through a contract with NASA, administered by the Oak Ridge Associated Universities. M.\,T. thanks Tryst DC Coffeehouse Bar and Lounge, where parts of this work were done, for its hospitality.

\bibliographystyle{physrev}
\bibliography{references}

\begin{thebibliography}{10}

\bibitem{Abadie:2010cfa}
LIGO Scientific, J.~Abadie {\em et~al.},
\newblock Class. Quantum Grav. {\bf 27}, 173001 (2010), arXiv:1003.2480.

\bibitem{Belczynski:2010tb}
K.~Belczynski {\em et~al.},
\newblock (2010), arXiv:1004.0386.

\bibitem{Kumar2005}
B.~V. K.~V. Kumar, A.~Mahalanobis, and R.~D. Juday,
\newblock {\em Correlation Pattern Recognition} (Cambridge University Press,
  2005).

\bibitem{Berti:2009kk}
E.~Berti, V.~Cardoso, and A.~O. Starinets,
\newblock Class. Quantum Grav. {\bf 26}, 163001 (2009), arXiv:0905.2975.

\bibitem{Abbott:2009km}
LIGO Scientific Collaboration, B.~Abbott {\em et~al.},
\newblock Phys. Rev. {\bf D80}, 062001 (2009), arXiv:0905.1654.

\bibitem{Berti:2007zu}
E.~Berti, J.~Cardoso, V.~Cardoso, and M.~Cavaglia,
\newblock Phys. Rev. {\bf D76}, 104044 (2007), arXiv:0707.1202.

\bibitem{Dreyer:2003bv}
O.~Dreyer {\em et~al.},
\newblock Class. Quantum Grav. {\bf 21}, 787 (2004), arXiv:gr-qc/0309007.

\bibitem{Berti:2005ys}
E.~Berti, V.~Cardoso, and C.~M. Will,
\newblock Phys. Rev. {\bf D73}, 064030 (2006), arXiv:gr-qc/0512160.

\bibitem{Kamaretsos:2011um}
I.~Kamaretsos, M.~Hannam, S.~Husa, and B.~Sathyaprakash,
\newblock (2011), arXiv:1107.0854.

\bibitem{Fregeau:2006yz}
J.~M. Fregeau, S.~L. Larson, M.~Miller, R.~W. O'Shaughnessy, and F.~A. Rasio,
\newblock Astrophys.J. {\bf 646}, L135 (2006), arXiv:astro-ph/0605732.

\bibitem{Berti:2008af}
E.~Berti and M.~Volonteri,
\newblock Astrophys.J. {\bf 684}, 822 (2008), arXiv:0802.0025,
\newblock * Brief entry *.

\bibitem{Owen:1995tm}
B.~J. Owen,
\newblock Phys. Rev. {\bf D53}, 6749 (1996), arXiv:gr-qc/9511032.

\bibitem{Creighton:1999pm}
J.~D. Creighton,
\newblock Phys. Rev. {\bf D60}, 022001 (1999), arXiv:gr-qc/9901084.

\bibitem{Nakano:2003ma}
H.~Nakano, H.~Takahashi, H.~Tagoshi, and M.~Sasaki,
\newblock Phys. Rev. {\bf D68}, 102003 (2003), arXiv:gr-qc/0306082.

\bibitem{Tsunesada:2004ft}
Y.~Tsunesada {\em et~al.},
\newblock Phys. Rev. {\bf D71}, 103005 (2005), arXiv:gr-qc/0410037.

\bibitem{Nakano:2004ib}
H.~Nakano, H.~Takahashi, H.~Tagoshi, and M.~Sasaki,
\newblock Prog.Theor.Phys. {\bf 111}, 781 (2004), arXiv:gr-qc/0403069.

\bibitem{Heng:2009zz}
I.~S. Heng,
\newblock Class. Quantum Grav. {\bf 26}, 105005 (2009).

\bibitem{Brady:2004cf}
P.~R. Brady and S.~Ray-Majumder,
\newblock Class. Quantum Grav. {\bf 21}, S1839 (2004), arXiv:gr-qc/0405036.

\bibitem{Cannon:2010qh}
K.~Cannon {\em et~al.},
\newblock Phys. Rev. {\bf D82}, 044025 (2010), arXiv:1005.0012.

\bibitem{Galley:2010rc}
C.~R. Galley, F.~Herrmann, J.~Silberholz, M.~Tiglio, and G.~Guerberoff,
\newblock Class. Quantum Grav. {\bf 27}, 245007 (2010), arXiv:1005.5560.

\bibitem{Field:2011mf}
S.~E. Field {\em et~al.},
\newblock Phys. Rev.Lett. {\bf 106}, 221102 (2011), arXiv:1101.3765.

\bibitem{Binev10convergencerates}
P.~Binev {\em et~al.},
\newblock SIAM J. Math. Analysis {\bf 43}, 1457 (2011).

\bibitem{Maggiore}
M.~Maggiore,
\newblock {\em Gravitational Waves Volume 1: Theory and Experiments} (Oxford
  University Press, Oxford, 2008).

\bibitem{Finn:1992wt}
L.~S. Finn,
\newblock Phys.Rev. {\bf D46}, 5236 (1992), arXiv:gr-qc/9209010.

\bibitem{Ajith:2009fz}
P.~Ajith and S.~Bose,
\newblock Phys.Rev. {\bf D79}, 084032 (2009), arXiv:0901.4936.

\bibitem{Messenger:2008ta}
C.~Messenger, R.~Prix, and M.~Papa,
\newblock Phys. Rev. {\bf D79}, 104017 (2009), arXiv:0809.5223.

\bibitem{Manca:2009xw}
G.~M. Manca and M.~Vallisneri,
\newblock Phys. Rev. {\bf D81}, 024004 (2010), arXiv:0909.0563.

\bibitem{Cormen:2001:IA:580470}
T.~H. Cormen, C.~Stein, R.~L. Rivest, and C.~E. Leiserson,
\newblock {\em Introduction to Algorithms}, 2nd ed. (McGraw-Hill Higher
  Education, 2001).

\bibitem{Salomon2010}
D.~Salomon and G.~Motta,
\newblock {\em Handbook of Data Compression}, 5th ed. (Springer-Verlag, 2010).

\bibitem{Quarteroni}
A.~Quarteroni, G.~Rozza, and A.~Manzoni,
\newblock Journal of Mathematics in Industry {\bf 1} (2011).

\bibitem{Cannon:2011tb}
K.~Cannon, C.~Hanna, D.~Keppel, and A.~C. Searle,
\newblock Phys. Rev. {\bf D83}, 084053 (2011), arXiv:1101.0584.

\bibitem{Cannon:2011vi}
K.~Cannon {\em et~al.},
\newblock (2011), arXiv:1107.2665.

\bibitem{LSD}
B.~Allen and {\it et al.},
\newblock \textsc LaL Software Documentation.

\bibitem{Cannon:2011xk}
K.~Cannon, C.~Hanna, and D.~Keppel,
\newblock (2011), arXiv:1101.4939.

\bibitem{RBM_Website}
\url{http://www.cscamm.umd.edu/people/faculty/tiglio/RB.html}.

\bibitem{Berti:2010ce}
E.~Berti {\em et~al.},
\newblock Phys. Rev. {\bf D81}, 104048 (2010), arXiv:1003.0812.

\end{thebibliography}

\end{document}